\documentclass[11pt]{article}

\usepackage[margin=1in]{geometry}
\usepackage[T1]{fontenc}
\usepackage[utf8]{inputenc}
\usepackage{lmodern}
\usepackage{natbib}
\usepackage{hyperref}
\usepackage{tcolorbox}
\usepackage{csquotes}
\usepackage{booktabs}
\usepackage{tabularx}
\usepackage{graphicx}
\usepackage{longtable} 
\DeclareUnicodeCharacter{2194}{$\leftrightarrow$}
\DeclareUnicodeCharacter{2212}{$-$}
\usepackage{booktabs}
\usepackage{tabularx}
\usepackage{array}
\usepackage{ragged2e}
\usepackage{authblk}

\newtcolorbox{refsbox}{%
  enhanced,
  breakable,
  colback=white,
  colframe=black!50,
  boxrule=0.6pt,
  arc=2pt,
  left=6pt,right=6pt,top=6pt,bottom=6pt
}

\title{The neuroscience of transformers}
\author[a,b]{Peter K\"onig}
\author[c]{Mario Negrello }

\affil[a]{Institute of Cognitive Science, University of Osnabr\"uck, 49090 Osnabrück, Germany}
\affil[b]{Department of Neurophysiology and Pathophysiology, University Medical Center Hamburg-Eppendorf, 20246 Hamburg, Germany}
\affil[c]{Department of Neuroscience, Erasmus MC, 3015 GE Rotterdam, The Netherlands}
\date{13/3/2026}

\begin{document}
\maketitle

\addcontentsline{toc}{section}{The neuroscience of transformers}

\tableofcontents
\newpage

\section{Abstract}
Neuroscience has long informed the development of artificial neural networks, but the success of modern architectures invites, in turn, the converse: can modern networks teach us lessons about brain function? Here, we examine the structure of the cortical column and propose that the transformer provides a natural computational analogy for multiple elements of cortical microcircuit organization. Rather than claiming a literal implementation of transformer equations in cortex, we develop a hypothetical mapping between transformer operations and laminar cortical features, using the analogy as an orienting framework for analysis and discussion. This mapping allows us to examine in greater depth how contextual selection, content routing, recurrent integration, and interlaminar transformations may be distributed across cortical circuitry. In doing so, we generate a broad set of predictions and experimentally testable hypotheses concerning laminar specialization, contextual modulation, dendritic integration, oscillatory coordination, and the effective connectivity of cortical columns. This proposal is intended as a structured hypothesis rather than a definitive account of cortical computation. Placing transformer operations and cortical architectonics into a common descriptive framework sharpens questions, reveals new functional correspondences, and opens a productive route for reciprocal exchange between systems neuroscience and modern AI. More broadly, this perspective suggests that comparing brains and architectures at the level of computational organization can yield genuine insight into both.

\section{Introduction}
\subsection{The brain and ANNs}

The development of artificial neural networks (ANNs) has, at least conceptually, been guided by the biological brain \citep{McCulloch1943A}. Early neural network models were explicitly framed as abstractions of cortical computation, and contemporary deep neural networks (DNNs)\citep{LeCun2015Deep} still borrow core ideas such as distributed representations and learning by adjusting synaptic weights \citep{Fukushima1980Neocognitr,Rosenblatt1958The,Rumelhart1986Learning}. In this sense, the brain has served as an existence proof that large networks of relatively simple units, trained with local updates, can support flexible and powerful cognition. Modern deep learning can be seen as a particularly successful instantiation of this program: by stacking multiple layers of simple, differentiable units and adjusting their connection strengths with gradient-based optimization\citep{LeCun1998Gradientba}, artificial systems now achieve performance that rivals or even exceeds human performance in certain domains\citep{Hassabis2017Neuroscien}.

At the same time, any close comparison between biological neural networks and standard ANNs reveals substantial differences. The cortex is densely and recurrently connected, with extensive feedback and lateral connections \citep{Douglas2004Neuronal,Harris2013Cortical}, whereas many of the most successful early deep architectures (e.g., convolutional networks for vision) were predominantly feedforward during inference \citep{Fukushima1980Neocognitr,Goodfellow2016Deep}. Biological neurons communicate via discrete spikes, operate under tight metabolic and temporal constraints, and exhibit complex dendritic integration \citep{Gerstner2014Neuronal}; standard ANN units compute continuous, deterministic activation functions, as is typical in mainstream deep learning formulations \citep{Goodfellow2016Deep}. The brain uses a wide variety of neurotransmitters and neuromodulators, implements learning across multiple timescales, and is embedded in a larger architecture of subcortical structures that shape learning, attention, and action selection \citep{Dayan2001Theoretica}. Even at the level of gross anatomy, the common analogy between “layers” in deep networks and cortical laminae, the 6-layered structure of the neocortex, is limited \citep{Mountcastle1997The}. Cortical laminae differ in their input--output connectivity, cell types, and biophysical properties\citep{Markram2006The}, yet the response properties of neurons within a given cortical column---across different laminae---can be surprisingly similar with respect to classical tuning measures (e.g., orientation selectivity in primary visual cortex) \citep{Hubel1962Receptive}. In contrast, artificial networks often implement layers as homogeneous collections of identical units, and functional differentiation across layers is an emergent property of the learned weights rather than reflecting distinct cell classes or canonical microcircuits \citep{Goodfellow2016Deep}. Thus, a multitude of differences between biological and artificial networks might have consequences that are little explored.

These differences are not accidental but the result of deliberate simplifications made for tractability and performance \citep{Marblestone2016Toward}. Deep learning focuses on a small subset of biologically inspired principles---hierarchical feature extraction, distributed coding, and supervised or self-supervised adaptation of synaptic weights---and discards many other aspects of brain organization that are difficult to model or train at scale \citep{LeCun2015Deep,Kriegeskorte2015Deep}. The success of DNNs in engineering tasks shows that this restricted subset of mechanisms is sufficient to support impressive capabilities, at least in constrained domains, and has in turn provided powerful models for interpreting neural data \citep{Yamins2016Using,Richards2019A}.

This became particularly evident in the context of visual object recognition. Deep convolutional networks trained on large-scale datasets such as ImageNet reached and then surpassed human-level performance in object classification \citep{Krizhevsky2012ImageNet,Russakovsky2015ImageNet}. The critical ingredients for this breakthrough were not more biologically realistic dynamics but rather the scaling of network size, the availability of massive labeled datasets, and efficient training procedures on modern hardware \citep{Krizhevsky2012ImageNet}. Increasing depth, width, and dataset size, together with architectural biases such as convolution and pooling, was sufficient to achieve state-of-the-art results \citep{Simonyan2015Very,He2016Deep,Huang2017Densely}. In this regime, the correspondence between biological and artificial networks was largely metaphorical (e.g., “layered processing similar to the visual hierarchy”) rather than a detailed mapping of neural circuitry, even though such models can be quantitatively compared to neural data in higher visual cortex \citep{Yamins2014Performanc,Khaligh-Razavi2014Deep}.

However, the fact that top-performing models omit many characteristics of real brains does not imply that these biological features are irrelevant to intelligence or to understanding neural computation. Recurrent connectivity, for example, is ubiquitous in the brain and underlies temporal integration, working memory, and context-sensitive processing \citep{Elman1990Finding,Hochreiter1997Long}; yet recurrence was long underutilized or heavily constrained in many high-performing vision models, despite evidence that recurrent convolutional networks can better capture aspects of biological object recognition \citep{Spoerer2017Recurrent}. Similarly, spiking dynamics \citep{Maass1997Networks,Gerstner2002Spiking}, diverse neurotransmitter systems and neuromodulation \citep{Marder2012Neuromodul}, and interactions with subcortical structures involved in learning and action selection (e.g., reward prediction signals) \citep{Schultz1997A} may be crucial for robustness, energy efficiency, online learning, or flexible behavior in natural environments---properties only partially captured by current ANNs. These aspects have been explored in specialized architectures (e.g., recurrent networks, spiking neural networks, or neuromodulation-inspired models), including work on deep learning with spiking neurons \citep{Pfeiffer2018Deep}, but they are not yet standard ingredients of the most widely used, top-performing systems.

Thus, in the era defined by ImageNet-scale benchmarks, the relationship between biological neural networks and mainstream artificial neural networks remained relatively loose \citep{Kriegeskorte2015Deep,Hassabis2017Neuroscien}. The brain inspired the initial conceptual framework and some coarse architectural motifs, but the details of cortical anatomy and physiology played little direct role in designing deep learning models that defined the state of the art \citep{Kriegeskorte2015Deep}. At the same time, these models proved useful as quantitative hypotheses about sensory representations \citep{Yamins2016Using} and helped motivate broader frameworks for how deep learning can interface with neuroscience \citep{Richards2019A,Hassabis2017Neuroscien}. This tension---between biological inspiration and engineering abstraction---provides an important backdrop for considering whether newer architectures, such as transformers (see next section), might bring artificial systems into closer alignment with our understanding of the brain, while also addressing limits of current deep learning paradigms \citep{Marcus2018Deep} and keeping in view the broader methodological role of models in network neuroscience \citep{Bassett2018On}.

\subsection{The revolution brought by the transformer architecture.}

The advent of the transformer architecture has fundamentally reshaped the landscape of artificial neural networks \citep{Vaswani2017Attention}. Introduced only a few years ago, transformers have rapidly become the dominant paradigm for sequence processing \citep{Devlin2019BERT,Brown2020Language} and, increasingly, for vision and multimodal integration \citep{Dosovitskiy2021An}. In many domains, they have delivered dramatic performance gains, enabling models that generate fluent text, translate between languages, solve complex reasoning problems, and integrate information across modalities at scales and levels of robustness previously unattainable \citep{Brown2020Language}. These advances have sparked intense discussion not only within machine learning and computational linguistics but also in the wider public sphere, where transformer-based models are increasingly perceived as a step change in artificial intelligence capabilities, including through the framing of such systems as “foundation models” \citep{Bommasani2021On}. In contrast to earlier deep learning breakthroughs around convolutional networks and ImageNet, which were largely confined to expert communities, transformer models have quickly become a focal point of societal debate, including prominent critiques of scale and downstream harms \citep{Bender2021On}.

Crucially, this leap in performance stems from a significant architectural change rather than merely scaling up existing designs. The transformer replaces the recurrent and convolutional backbones that dominated earlier sequence models \citep{Bahdanau2015Neural} with a self-attention mechanism \citep{Vaswani2017Attention}. A defining feature of self-attention is a multiplicative interaction among different parts of the input: each input element (e.g., a token in a sentence) generates query, key, and value representations, and the relevance of one element to another is computed via dot products between queries and keys \citep{Vaswani2017Attention}. These multiplicative similarity scores are then used to weight the contributions of the corresponding values. In other words, the model explicitly computes how much “attention” each element should pay to every other element and then integrates information by applying these learned, context-dependent weights \citep{Vaswani2017Attention}. Stacked across multiple layers and heads, this yields a flexible mechanism for contextualizing each representation in light of the entire input sequence, while also motivating work on more efficient attention variants \citep{Choromanski2021Rethinking} and enabling straightforward extensions of the same core mechanism beyond language, including vision \citep{Dosovitskiy2021An}.

From the perspective of neural computation, this is a striking development. Multiplicative interactions---whether in the form of gain modulation, gating, or context-dependent weighting---have a long but intermittent history in theoretical and systems neuroscience \citep{Salinas2000Gain}. Models of attention \citep{Treue1999Featurebas,Reynolds2009The}, coordinate transformations \citep{Pouget1997Spatial}, and mixed selectivity \citep{Rigotti2013The} have invoked multiplicative or bilinear terms to explain how neurons combine different sources of information. However, these ideas were typically explored in small-scale, task-specific models and did not crystallize into a widely adopted architectural principle for large, trainable networks. Moreover, the specific structure of transformer-style self-attention---with its separation into queries, keys, and values, its explicit all-to-all interaction pattern, and its central role in high-capacity sequence models---has not been a major focus of prior neuroscientific discussion. In this sense, the transformer architecture introduces a powerful new computational motif into mainstream artificial networks that does not map straightforwardly onto the familiar list of brain--model mismatches (e.g., spiking vs. rate units, recurrence vs. feedforward, cortical vs. subcortical structures) outlined for earlier deep networks. It is not simply a more “brain-like” CNN or RNN; it embodies a qualitatively different way of structuring interactions within the network.

One consequence of this development is that research in computational neuroscience and in state-of-the-art artificial neural networks has, at present, become largely decoupled \citep{Hassabis2017Neuroscien,Richards2019A}. While there is growing interest in using modern deep networks, including transformers, as descriptive models of behavior and brain data \citep{Kriegeskorte2015Deep,Yamins2016Using}, and recent work has reported partial convergence between language models and neural/behavioral measures in NLP settings \citep{Schrimpf2020Artificial,Caucheteux2022Brains}, few neurobiologists treat the transformer architecture itself as a template for understanding cortical or subcortical computation. The dominant theoretical frameworks in systems neuroscience still emphasize recurrent dynamics, attractor states, probabilistic population codes, and canonical microcircuits, rather than attention-like multiplicative routing mechanisms implemented at the scale and form seen in transformers. Conversely, many machine learning researchers who work on transformer-based models do not look to the brain for architectural inspiration in the way earlier generations of deep learning occasionally did \citep{Hassabis2017Neuroscien}. Progress is often driven by empirical scaling, optimization tricks, and engineering constraints, with biological considerations playing at most a peripheral role.

This situation is somewhat paradoxical. At the very moment artificial systems have achieved unprecedented performance across a wide range of tasks, direct comparisons between these systems and the brain have receded from the forefront of both communities. This occurs despite prominent arguments that modern deep learning can be understood as a form of “direct fit” to natural data \citep{Hasson2020Direct} and ongoing debates about what deep learning does and does not explain \citep{Saxe2021If,Marcus2018Deep}. The transformer architecture has been embraced as an extraordinarily effective tool and has helped motivate the framing of large-scale models as “foundation models” \citep{Bommasani2021On}, yet its relationship to known neural mechanisms remains underexplored. Likewise, neuroscience has accumulated a rich body of work on multiplicative interactions, attention, and context-dependent modulation, but has not yet systematically engaged with the specific computational schema that has proven so powerful in artificial models. This growing disconnect raises substantive questions: Are transformers implementing a computational principle that the brain also exploits, but in a different guise? Or do they reflect a fundamentally non-biological solution that is well suited to digital hardware and large datasets, but only loosely related to the mechanisms of biological cognition? In the following, we will argue that closing this conceptual gap---by analyzing transformers through the lens of neuroscience and vice versa---may be essential for understanding these models and for refining our theories of brain function, while keeping in view the broader methodological role (and limits) of models in network neuroscience \citep{Bassett2018On}.

\section{Transformers in the brain}


\begin{table*}[t]
\centering
\caption{\textbf{Mapping transformer components onto cortical mechanisms (columns as modules).}}
\label{tab1:transformer_cortex_mapping}
\footnotesize1
\setlength{\tabcolsep}{4pt}        
\renewcommand{\arraystretch}{1.05} 

\begin{tabular}{@{}p{0.22\textwidth}p{0.76\textwidth}@{}}
\toprule
\textbf{Transformer concept} & \textbf{Brain analogue (cortex-centric)} \\
\midrule
Token & \textbf{A cortical column (module)}; columnar structure laid out over a topographic map form the “set of tokens” that interact \\
\midrule
Input embedding & \textbf{Thalamic / subcortical drive} (e.g., LGN$\rightarrow$V1), providing a shared “input embedding” to the local circuit \\
\midrule
Values (V) & L4$\to$L2/3 projection stream, the canonical granule-cell-to-supragranular feedforward pathway. \\
\midrule
Queries (Q) & \textbf{L1 contextual activity} long-range feedback projections carrying top-down task goals and predictions that
determine which incoming values are retrieved.\\
\midrule
Keys (K) & \textbf{Tangential stream in laminae 2/3 and 5} \\
\midrule
Attention weights (soft selection) & Multiplicative gates Q$\cdot$K; \textbf{gain modulation + dendritic nonlinearities} as multiplicative gates; long-range horizontal/feedback structure implements “who attends to whom” \\
\midrule
Self-attention connectivity pattern & All-to-all (or sparse) token–token interaction; \textbf{horizontal (tangential) + feedback connectivity between columns}, implementing relations across positions/features (including grouping-like links) \\
\midrule
Multi-head attention & Many parallel routing schemes over the same input; \textbf{multiple neuronal subpopulations / dynamical patterns within a column} sharing the same drive but sampling intracortical context differently (“heads”) \\
\midrule
FFN / MLP sublayer & Local per-token recoding (often expand$\rightarrow$nonlinearity$\rightarrow$compress); \textbf{local recoding across laminar microcircuits} remapping into a new feature space (“feed-forward recoding”) \\
\midrule
Residual pathway & Add prior representation with update; \textbf{pooling / mixing} with ongoing intracortical + top-down influences \\
\midrule
LayerNorm (“Norm”) & Stabilize scale; aid optimization and compositionality; \textbf{normalization-like mechanisms} implied via gain control/normalization motifs accompanying attention-like gating (functional, not literal operator) \\
\midrule
Stack of blocks & Depth via serial composition (feedforward per pass); \textbf{hierarchically stacked cortical modules} with abundant \textbf{feedback}; higher modules shape interaction structure in lower modules \\
\bottomrule
\end{tabular}
\end{table*}

\begin{table*}[t]
\centering
\caption{\textbf{Relaxed assumptions.}}
\label{tab2:relaxed_assumptions}
\footnotesize
\setlength{\tabcolsep}{4pt}
\renewcommand{\arraystretch}{1.05}

\begin{tabular}{@{}p{0.18\textwidth}p{0.32\textwidth}p{0.46\textwidth}@{}}
\toprule
\textbf{Theme} & \textbf{Transformer default} & \textbf{Cortex} \\
\midrule
Literal equation-level implementation & Exact Q,K,V matrices + softmax attention computed as specified & Cortex implements the \textbf{same computational motif} (context-dependent multiplicative routing), \textbf{not} the literal equations “token by token” \\
\midrule
Temporal unfolding & A forward pass is essentially feedforward and static during inference & Cortical columns are \textbf{strongly recurrent} and evolve over time; laminar loops may act as a static transform on a task-relevant timescale \\
\midrule
No top-down influence during a pass & Standard transformer block does not receive explicit top-down cortical-style feedback during the same pass & Cortex has ubiquitous \textbf{feedback}; top-down + lateral projections extend the module beyond the canonical transformer diagram \\
\midrule
Token granularity & Tokens are discrete symbols/patches chosen by a tokenizer/patchifier & Tokens are \textbf{features} in a map (continuous, biophysical substrates) interacting via long-range connectivity \\
\midrule
Input dominance & Input embedding is a major driver of hidden state & Thalamic drive can be a \textbf{minor fraction} of synapses (e.g., 10–15\% onto recipient neurons), with computation dominated by intracortical context \\
\midrule
Explicit heads as separate parameter blocks & Heads are explicit parallel projections with clean vector-space semantics & Heads correspond to \textbf{subpopulations} sharing inputs but differing by connectivity and context sampling \\
\midrule
Attention as an explicit weight matrix & Attention weights are an explicit, separately computed matrix per head & Gating emerges from \textbf{gain modulation + dendritic nonlinearities} (interacting via horizontal and feedback connectivity), not necessarily an explicit “attention matrix” \\
\midrule
Single step update & Each layer computes a defined transformation once per step & Cortex supports \textbf{iterative refinement} via recurrent/feedback loops; “one box” corresponds to a transformer-like operation embedded in ongoing cycles \\
\midrule
Purely local FFN meaning & FFN is a per-token MLP with no explicit anatomical substrate & “FFN” is mapped to \textbf{local laminar recoding} and interlaminar transfer; also influenced by top-down integration timing \\
\bottomrule
\end{tabular}
\end{table*}

In this section, we move beyond this paradox and outline a concrete, albeit speculative, hypothesis about the relationship between transformers and the brain. Rather than treating transformers as intrinsically “non-biological,” we propose that key aspects of the architecture can be understood as a particular way of organizing computations that the mammalian cortex may already approximate, especially given the long-standing role of multiplicative gain control and normalization in neural computation \citep{Salinas2000Gain,Reynolds2009The}. The central move is to reinterpret the basic unit of comparison: instead of mapping an entire cortical area onto a single “layer” of an artificial network, we treat a cortical column and its laminar microcircuit as a functional module, with mappings as in \ref{tab1:transformer_cortex_mapping}. We then ask whether such a module could implement the same kind of context-dependent, gated interactions over shared inputs that define a transformer block \citep{Vaswani2017Attention}.

More specifically, we will consider whether the internal circuitry of a column can be mapped onto the core components of a transformer block---queries, keys, and values linked by multi-head attention, followed by a local feed-forward recoding \citep{Vaswani2017Attention}. On this view, laminar pathways and recurrent loops within a column implement functionally analogous processes to self-attention, while the column’s local input--output transformation plays the role of the feed-forward sublayer. We do not claim that this mapping is unique, anatomically literal, or complete; it is offered as an existence proof and organizing hypothesis. The aim is to show that once we choose an appropriate level of abstraction, the laminar structure of the cortical column can be viewed as instantiating a transformer-like computational motif, in the same spirit as broader efforts to connect modern deep learning models and neuroscience \citep{Richards2019A} and as proposals emphasizing rich, context-dependent neural representations such as mixed selectivity \citep{Rigotti2013The}.

\subsection{Elements of the mapping}

A starting point is to reconsider the standard mapping between brains and artificial networks. In most comparisons, an entire cortical area is loosely equated with a single “layer” in a deep network \citep{LeCun2015Deep,Felleman1991Distribute}, and the rich laminar structure within that area is collapsed into a single homogeneous set of units, despite extensive evidence for systematic laminar and circuit-level organization in cortex \citep{Mountcastle1997The,Douglas2004Neuronal,Harris2015The}. All neurons within a cortical column---across laminae 2/3, 4, 5, and 6---are, in effect, reduced to a single ANN unit or a uniform population within one artificial layer. This is a drastic simplification. Here, we propose that this simplification misses an essential aspect of cortical circuitry and hinders understanding transformer-like computation in the cortex. Instead, what we usually call a “layer” in an ANN should be thought of as corresponding to a module built from several cortical laminae and their recurrent loops, not necessarily to a single anatomical lamina or to a whole area. In this view, the laminar circuit within a cortical column is not an implementation detail that can be ignored but a substrate that realizes a transformer-like operation.

Figure 1 develops this idea step by step. Panel 1A reproduces, in simplified form, the original encoder--decoder transformer from “Attention is all you need” \citep{Vaswani2017Attention}: on the encoder side, an input embedding is processed by a stack of blocks, each consisting of multi-head attention followed by Add \& Norm, then a feed-forward sublayer followed again by Add \& Norm. On the decoder side, an output embedding passes through masked multi-head attention and Add \& Norm, then a second multi-head attention (attending to the encoder output) with Add \& Norm, and finally a feed-forward sublayer with Add \& Norm. In Panel 1B, this architecture is redrawn without changing its computations. The encoder and decoder blocks are rearranged and grouped so that the sequence “multi-head attention → Add \& Norm → feed-forward → Add \& Norm” appears as a recognizable processing unit, and the corresponding decoder sequence “masked multi-head attention → Add \& Norm → multi-head attention → Add \& Norm → feed-forward → Add \& Norm” is laid out to make its internal structure more transparent. This kind of schematic reorganization is also useful for relating the original sequence-to-sequence transformer to later encoder-only variants used in language modeling \citep{Devlin2019BERT} and to the broader family of transformer architectures now deployed well beyond NLP, including in vision \citep{Khan2022Transforme}. At this stage, no new components are introduced; the circuit is only reorganized.

Panel 1C makes explicit a structural feature that is implicit in the canonical diagram. For our purposes, we introduce the term module to denote a self-contained subcircuit that (i) establishes an input embedding, (ii) applies its own multi-head attention to that embedding, (iii) passes the result through a local feed-forward transformation, and (iv) exposes a clearly defined output embedding, with residual and normalization operations wrapped around these stages \citep{Vaswani2017Attention}. A module in this sense is designed to be reusable: it can be instantiated many times in parallel (for example, across a topographic map) or stacked in depth (for hierarchical processing), without requiring additional, bespoke interconnection machinery beyond its standard input--output interfaces.

When we examine the original encoder--decoder architecture of “Attention is all you need” through this lens \citep{Vaswani2017Attention}, we find that some functionally paired operations---for example, a particular attention transformation and the feed-forward recoding that naturally follows it---are distributed across neighboring blocks when we try to align them with repeating, reusable units and a single laminar microcircuit per unit. In that sense, the canonical diagram effectively spans about one and a half of our modules. In Panel 1C, we “complete” this pattern to obtain two full, symmetric modules by adding the missing counterparts, so that each module now contains its own attention and feed-forward stages, together with their associated residual and normalization operations. These two modules can then be stacked serially (as in a depth hierarchy) or arranged in parallel pathways, in a way that parallels the general role of residual pathways in stabilizing deep architectures \citep{He2016Deep}. Importantly, this completion does not alter what the overall transformer computes; it simply carves the existing computation into two self-contained, transformer-like units that are easier to compare to a repeating biological microcircuit. This modular view is also consistent with the idea that transformer feed-forward sublayers can be interpreted as learned memory-like components \citep{Geva2021Transforme}.

Now we take the crucial, and admittedly speculative, step of mapping the elements of one such module onto the laminae of a cortical column (Fig. 1D and E). Here, different cortical laminae are aligned with different parts of the module: external signals targeting lamina 4 are subject to an input embedding and then provide “values” towards the supragranular laminae. Activity in laminae 2/3 and 5 shapes “queries” and “keys” through recurrent and feedback connections within canonical cortical circuitry \citep{Douglas2004Neuronal,Harris2013Cortical}; and local gain modulation and dendritic nonlinearities act as gates that weight these inputs multiplicatively, analogous to attention weights \citep{Salinas2000Gain,Reynolds2009The,Larkum2013A}. The dashed arrows in the figure indicate additional top-down and tangential (horizontal) projections that are not explicitly present in the original transformer diagram but connect different transformer blocks. These connections are ubiquitous in the cerebral cortex and naturally extend the module to include feedback and lateral modulation, in line with common “canonical microcircuit” perspectives and predictive-coding-inspired accounts of laminar message passing \citep{Bastos2012}. The mapping of individual subcomponents to specific laminae is explicitly labeled as speculative in the figure, and we have not fundamentally changed the underlying transformer computations---only rearranged and slightly augmented them to align with a plausible laminar circuit.

Our claim is not that the cortex literally implements the transformer equations token by token, but that the same computational pattern---context-dependent, multiplicative routing over shared inputs, realized within a layered, recurrent circuit---may underlie both systems, consistent with the broader view that modern deep networks can serve as useful (if imperfect) computational hypotheses about brain function \citep{Kriegeskorte2015Deep}. By reassigning the correspondence between ANN “layers” and cortical modules and interpreting laminar interactions as implementing attention-like gating, the figure shows how a standard transformer can be mapped, with minor rearrangement, onto the microcircuit of a cortical column. In this mapping, a module corresponds to one such column and its laminar circuitry; it receives a shared input embedding (for example, thalamic drive), applies context-dependent weighting of that input via attention-like interactions, and produces an updated output embedding that is passed on to other modules.

Across the cortex, many of these modules are arranged in parallel, for instance, over retinotopic or somatotopic maps. In the transformer analogy, these modules collectively play the role of the tokens over which self-attention operates: different columns encode different positions or feature combinations, and long-range horizontal and feedback connections implement the “who attends to whom” relationships across this population. Within each module, multiple neuronal subpopulations and dynamical patterns can be viewed as implementing different attention “heads,” all drawing on similar external inputs but sampling and combining intracortical context in distinct ways. Thus, attention in our proposal is inherently multi-scale: it is realized both within a column, through laminar loops and gain modulation, and between columns, through tangential and feedback connectivity that links different positions and features. Framed in this way, the laminated structure of the cortical column becomes a plausible instantiation of a transformer-like computational motif, and this, in turn, motivates a more systematic exploration of transformers as candidate models of cortical computation and of cortical architecture as a source of constraints and inspiration for future transformer-like designs, in line with broader proposals to use deep networks as computational hypotheses for neuroscience \citep{Kriegeskorte2015Deep} while also engaging with ongoing debates about what exactly contemporary deep learning explains about cognition and biology \citep{Hasson2020Direct,Saxe2021If} and with frameworks for a productive deep learning--neuroscience interface \citep{Richards2019A}.

It is important to emphasize that this mapping is intended at the level of effective input--output transformations, not as a claim about the detailed temporal unfolding of cortical dynamics. In contrast to the essentially feedforward computation of a transformer block during a single forward pass, cortical columns are embedded in strongly recurrent circuits whose activity evolves over time. Here, we adopt the simplifying perspective that, on the timescale relevant to a given computation, the recurrent laminar loops within a column can be approximated by a quasi-static transformation with attention-like gating; a fuller treatment of the temporal dynamics and their algorithmic role lies beyond the scope of this article.

\subsection{A detailed walkthrough}

In this section, we walk through the circuit diagram in Figure 1D,E,F step by step and relate each component to known features of cortical microcircuitry \citep{Douglas2004Neuronal,Harris2015The}. For clarity, we treat each “box” in the figure as representing the neuronal circuitry of a single cortical column, including its laminar structure and local recurrent connections \citep{Mountcastle1997The}. As argued above, the standard transformer architecture is effectively distributed across one and a half of these boxes; here, however, we focus on a single box and describe its components using terminology natural for both transformers (queries, keys, values, feed-forward stages) and cortical columns (granular, supragranular, and infragranular laminae), drawing where useful on canonical-microcircuit and predictive-coding perspectives on laminar message passing \citep{Bastos2012}.

\begin{figure}[p]
  \centering
    \includegraphics[width=\textwidth,height=0.95\textheight,keepaspectratio]{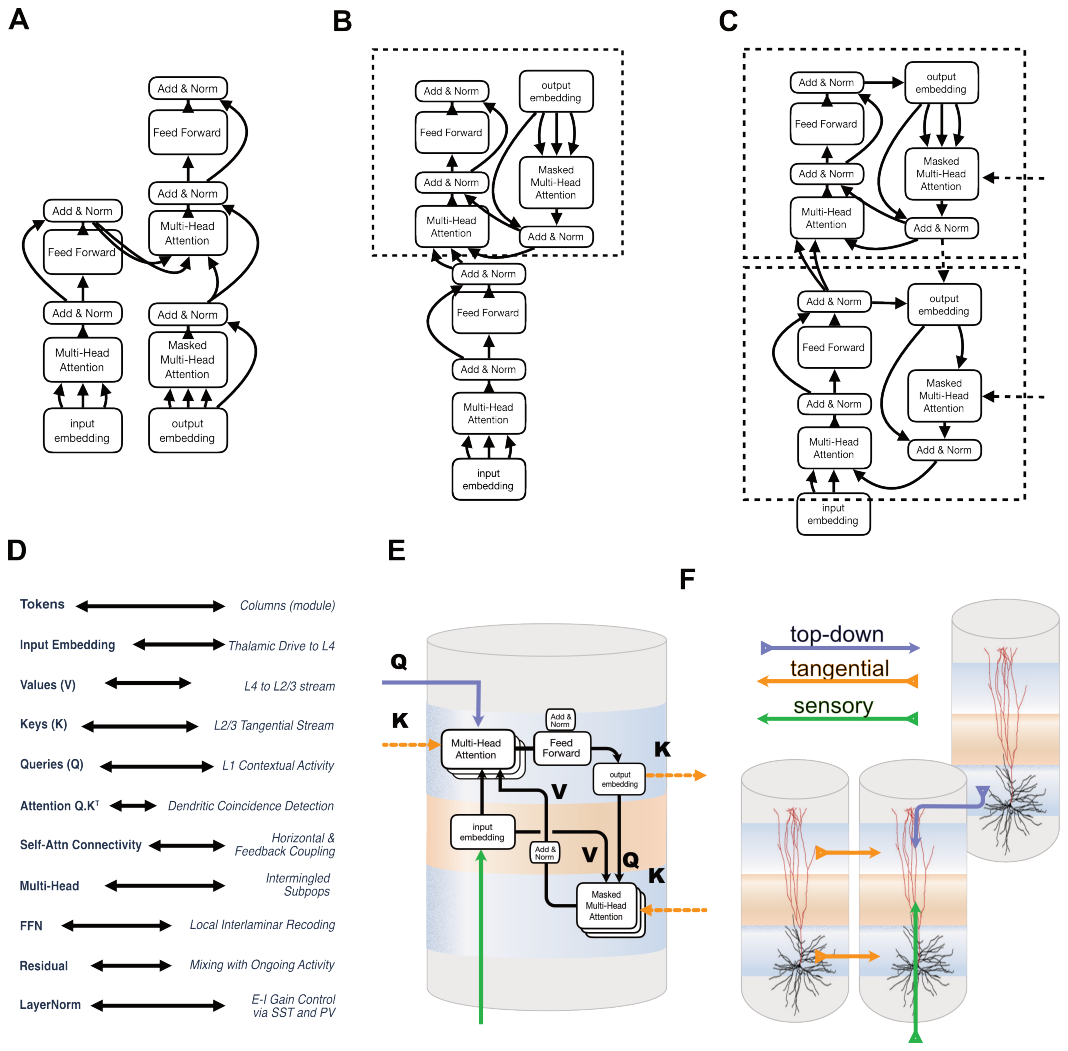}

  \caption{\textbf{Mapping the transformer encoder/decoder architecture onto the laminar structure of the cortical column.}}
  \label{fig:main}
\end{figure}
{\footnotesize
\noindent\textbf{Figure 1. Mapping the transformer encoder/decoder architecture onto
the laminar structure of the cortical column.}

\textbf{(A)} Standard transformer architecture \citep{Vaswani2017Attention}, comprising an encoder stack (Multi-Head Attention $\to$ Add \& Norm $\to$ Feed Forward $\to$ Add \& Norm) and a decoder stack (Masked Multi-Head Attention $\to$ Add \& Norm $\to$ cross-attention $\to$ Add \& Norm $\to$ Feed Forward $\to$ Add \& Norm $\to$ output embedding). The decoder is conditioned on encoder output via cross-attention.

\textbf{(B)} When the decoder stack is inverted vertically, its processing sequence aligns naturally with the laminar depth axis of cortex: deep layers (L5/6) correspond to the decoder-like autoregressive component, whereas superficial layers (L2/3) correspond to the encoder-like component. The dashed boundary marks the operations proposed to occur within a single cortical column treated as a combined encoder/decoder module.

\textbf{(C)} The single-column module from (B) shown as a recurrent, stackable unit. Repetition of this module across areas would correspond to hierarchical unrolling of the transformer (e.g.\ V1$\to$V2$\to$V4$\to$IT), with feedforward projections carrying encoder outputs upward and feedback projections carrying decoder/context signals downward. The recurrent arrow denotes the masked/autoregressive input derived from the module’s own prior output.

\textbf{(D)} Proposed correspondence between transformer components and cortical circuitry. \textit{Tokens} correspond to cortical columns. \textit{Input embedding} corresponds to thalamic drive to L4. \textit{Values} (V) correspond to the canonical L4$\to$L2/3 feedforward pathway. \textit{Keys} (K) correspond to tangential L2/3 interactions mediated by horizontal collaterals of supragranular pyramidal neurons. \textit{Queries} (Q) correspond to contextual feedback terminating in L1 on distal apical tufts of L2/3 and L5 pyramidal neurons. \textit{Attention} ($Q\!\cdot\!K^\top$) corresponds to dendritic coincidence nonlinearities, such as supralinear calcium events driven by coincident apical and basal/perisomatic input. \textit{Self-attention connectivity} corresponds to horizontal and feedback coupling within and between areas. \textit{Multi-head attention} corresponds to parallel functional subpopulations, either across neighbouring columns or intermingled within a column. \textit{FFN} corresponds to local interlaminar recoding within the column. \textit{Residual} corresponds to persistent recurrent activity maintaining an ongoing representational state. \textit{LayerNorm} corresponds to E--I gain control mediated by inhibitory interneuron circuitry.

\textbf{(E)} Proposed laminar circuit implementing the combined encoder/decoder computation within a single cortical column. Sensory drive ascends from thalamus to L4, forming the input embedding and supplying Values ($V$) to the L2/3 encoder-like stage. Keys ($K$) are conveyed tangentially through L2/3 horizontal projections, broadcasting the column’s current representational state to neighbouring columns while receiving theirs. Queries arrive as contextual input to L1 apical dendrites. Their interaction in supragranular pyramidal neurons implements an attention-like matching operation, after which the representation passes through gain normalisation and local feed-forward recoding. The resulting encoded representation is then passed to L5, where recurrent state supplies the autoregressive input required for decoder-like masked attention. The L5 output is then broadcast laterally and propagated onward through the cortical hierarchy.

\textbf{(F)} Inter-columnar interactions proposed to implement self-attention and multi-head-like computation across the cortical sheet. Top-down signals arriving in L1 provide a shared query broadcast to multiple columns. Tangential L2/3 connections distribute keys laterally according to the structured patchy connectivity of cortex, linking columns with related feature preferences. Thalamic input to L4 provides column-specific values. Parallel processing across multiple columns, or across distinct functional subpopulations within a column, is proposed to realise multiple attention heads operating simultaneously over the same sensory input.

We begin with the subcortical input. In the primate visual cortex, the projection from the lateral geniculate nucleus (LGN) of the thalamus is considered to be the critical input. These afferents primarily target neurons in lamina 4 and, to a lesser extent, those in lamina 6. Anatomical studies have consistently shown that this thalamic drive constitutes only a small fraction of all synapses onto the recipient cortical neurons, with estimates of 10--15\% of the total input each neuron receives \citep{Peters1993Numerical,Ahmed1994Polyneuron}. From an engineering perspective, this is striking: the very neurons that receive the “external” sensory input are, in terms of synapse count, dominated by recurrent input from within the cortex \citep{Binzegger2004A,Douglas2007Recurrent,Martin2014Microcircu}. We interpret the processing of the thalamo-cortical input in layer 4 of primary visual cortex as constructing an input embedding (Fig. 1E), where the bulk of the computation arises from internally generated interactions that evaluate and re-weight this afferent signals.

The supragranular laminae implement the next step, a multi-head attention block. Its functional realization varies across the cortical hierarchy. In early sensory cortices, the high predictability of natural stimulus statistics—governed by Gestalt principles such as collinearity and contour continuation \citep{Betsch2004The}—obviates the need for the continuous, flexible Query-Key alignment required in higher association areas. Instead, these regions likely employ an "attention-free" transformer variant, where routing probabilities are structurally hardwired into the long-range tangential connectivity \citep{Ko2011Functional}. In this framework, the synaptic weight matrix of these tangential paths effectively serves as a fixed attention matrix. The necessary multiplicative interaction is then realized through temporal coincidence: gamma-band oscillations induce phase-locked synchronization between compatible columns \citep{Gray1989Oscillatory, Singer1999Neuronal, Fries2015Rhythms}, acting as a dynamic gate that binds distributed features and routes their combined Values downstream. By marrying static structural matrices with oscillatory coincidence detection, early sensory hierarchies can achieve rapid, multi-head spatial routing without the computational overhead of dynamic dot-product similarity.

The architectural hallmark of modern transformers—Multi-Head Attention (MHA)—finds a robust biological analogue in the fine-scale functional and structural diversity of neurons within a single cortical column. While classical models often view the column as a monolithic module of units with identical tuning, detailed physiological mappings reveal substantial scatter in neuronal response properties \citep{DeAngelis1999Functional, Ringach2002Orientatio}. Under our framework, this "functional scatter" provides the anatomical substrate for parallel attention heads: whereas nearby neurons share a common "Value" embedding via overlapping subcortical input—manifesting as similar classical receptive fields \citep{Hubel1962Receptive, Hubel1968Receptive}—they diverge sharply in their non-classical receptive fields, which are driven by horizontal and feedback connections \citep{Das1999Topography, Fitzpatrick2000Seeing}. This allows distinct neuronal subpopulations within the same column to generate unique Queries and Keys by sampling intracortical context through specific recurrent and long-range lateral connections \citep{Gilbert1983Clustered}. Consequently, the column does not compute a single monolithic match; rather, it executes a high-dimensional routing operation where relatively compressed, low-rank Query-Key spaces govern the selective transmission of higher-dimensional Value representations. This predicts that while columnar responses may appear redundant under simple stimuli, they will diverge during complex tasks as individual "heads" engage with distinct relational structures across the stimulus ensemble \citep{Cossell2015Functional}.

After the attention-like stage, activity is transformed and recoded by the equivalent of the transformer feed-forward sublayer. In the model, this stage applies a position-wise non-linear transformation to each element, often expanding and then compressing the representation. In the cortical column, we argue that this recoding occurs as activity is further processed in the supra-granular laminae. They remap the representation into a new feature space, supported by canonical neocortical microcircuitry and intralaminar connectivity \citep{Douglas2004Neuronal,Thomson2003Interlamin}. The output of this recoding stage---the analogue of the transformer’s output embedding---receives top-down projections from higher-order areas, consistent with known hierarchical feedforward and feedback pathways in primate visual cortex \citep{Markov2014Anatomy}. These top-down signals are integrated into a joint representation only after the local computation in the box is carried out. At the same time, the “pure” result of the local computation is already passed forward to the next box (i.e., to the next cortical area or module) before full integration of top-down context. This sequencing mirrors the idea that the cortex can propagate a rapidly computed feedforward estimate while still refining the local representation through slower recurrent and feedback interactions \citep{Lamme2000The}, and it aligns with evidence for layer-specific functional roles during contextual computations such as figure--ground segregation in V1 \citep{Self2013Distinct}.

As the next step, the activity pattern generated in the output embedding, influences the infragranular laminae, which in turn project to subcortical targets and other cortical areas. There are at least two major pathways for this transfer. One is the direct projection from supragranular pyramidal neurons to neurons in laminae 5 and 6, mediated by axonal connections; the other is the propagation of signals along the apical dendritic trees of large laminae 5 pyramidal neurons, which extend into upper laminae and can integrate distal inputs there \citep{Thomson2003Interlamin}. A complication of our mapping is that much of the tangential (horizontal) connectivity is concentrated in supragranular laminae, including prominent intrinsic/lateral networks in primate and cat visual cortex \citep{Rockland1983Intrinsic,Gilbert1983Clustered,Stettler2002Lateral}. In Figure 1D, the projection from these laminae to the “output embedding” can be seen as analogous to the Add \& Norm step that pools local activity, including short-range tangential inputs that are relatively unspecific to fine-grained receptive field properties. This pooling allows for substantial recoding of the representation that need not preserve the original selectivity in a one-to-one fashion. By contrast, the key--query interactions that define which inputs should be bound together in a context-specific way may be more strongly associated with long-range horizontal and feedback projections, implementing something like Gestalt grouping principles---linking elements that belong to the same contour, object, or pattern across space and features---in line with work on feedforward, lateral, and feedback contributions to classical and extra-classical receptive field structure in V1 \citep{Angelucci2006Contributi}.

The apical dendrites of lamina 5 pyramidal neurons deserve special mention. Physiological and modeling work has highlighted their role as potential “gating” or “mode-switching” elements, capable of integrating distal inputs from higher-order areas and modulating the influence of basal, feedforward inputs on spike output \citep{Larkum1999A,Larkum2004Topdown,Major2013Active}. This has led to proposals that apical dendrites implement context- or error-dependent modulation of local processing \citep{Körding2000Learning,Urbanczik2014Learning,Larkum2013A}, and even theories that tie their dynamics to different phases of learning across wakefulness and sleep. Similar ideas about the special role of apical function in contextual modulation and cognitive coordination have been developed by Phillips and colleagues \citep{Phillips2018Apical,Phillips2013Convergenc}. These ideas are deeply compatible with our general emphasis on multiplicative, context-dependent interactions and may provide important biological mechanisms for implementing aspects of attention and credit assignment. At the same time, we treat them here as partially orthogonal to our central claim. The present hypothesis does not depend on any single, specific dendritic mechanism; rather, it relies on the broader laminar and recurrent organization of the column. Detailed apical computations may refine or enrich the mapping between transformers and cortical microcircuits, but they are not essential for the basic correspondence illustrated in Figure 1E. More generally, related normative and learning-based accounts of how receptive field-like invariances can be adapted from natural stimulus statistics provide a complementary perspective on how such dendritic/laminar mechanisms might be shaped by experience \citep{Körding2001Generation,Einhäuser2002Learning,Körding2004How}. Also note, that the “query--key space” accessed in supragranular and infragranular laminae differ: different sets of local / long-range inputs and static / dynamic attention mechanisms shape which relational structures are emphasized.

Finally, the circuit closes. Activity shaped by the local self-attention-like interactions, recoded by feedforward-like transformations, and integrated with top-down context is routed onward via long-range projections to other cortical columns and subcortical structures, within a broader distributed hierarchical organization of cortex \citep{Felleman1991Distribute,Hilgetag2020Hierarchy}. At the same time, feedback from those downstream targets returns to the original column, influencing the next processing cycle, consistent with known feedforward and feedback pathways and their distinct functional signatures \citep{Markov2014Anatomy,Bastos2015}. In the language of transformers, a single pass through one box implements a transformer-like operation on the current pattern of activity; stacking and interconnecting many such boxes yields a recurrent, distributed network that, in principle, can approximate the behavior of a deep transformer stack. Thus, by tracing the flow of activity around this loop, we see how the proposed correspondence between transformer components and cortical laminae gives rise to a coherent, closed processing cycle within and between cortical columns, grounded in cortical connectivity constraints and sensory coding considerations \citep{Harris2013Cortical}.

The hypothesis is deliberately mechanistic enough to yield concrete, testable predictions, summarized in tables \ref{tab:appendix-a1} and \ref{tab:appendix-a2}. First, if multiple attention-like “heads” are instantiated by different neurons or subpopulations within a column that share a common input embedding but differ in how they sample intracortical context, then their response properties should remain relatively similar under simple, feedforward-dominated stimulus conditions but diverge systematically when rich contextual structure or complex tasks are engaged. In particular, highly controlled receptive-field mapping with simple stimuli should underestimate the diversity of tuning that becomes apparent under naturalistic or task-dependent conditions within the same column.


\renewcommand{\arraystretch}{1.15}
\setlength{\tabcolsep}{4pt}
\small

\begin{longtable}{p{0.16\textwidth} p{0.18\textwidth} p{0.24\textwidth} p{0.34\textwidth}}
\caption{Detailed mapping with testable predictions.}\label{tab:appendix-a1}\\
\toprule
\textbf{Transformer concept} & \textbf{In-transformer role} & \textbf{Proposed brain analogue} & \textbf{Testable predictions} \\
\midrule
\endfirsthead
\toprule
\textbf{Transformer concept} & \textbf{In-transformer role} & \textbf{Proposed brain analogue} & \textbf{Testable predictions} \\
\midrule
\endhead
\bottomrule
\endfoot
Token & Unit participating in attention & Column/module within a topographic sheet & Effective coupling between nearby/distant columns should shift with task context; perturbing one column should change context-dependent influence patterns across the sheet. \\
Input embedding & Shared representational drive & Thalamic/subcortical input & Trial-by-trial variability in thalamic drive should explain less variance than intracortical context in many regimes; stronger context domination during ambiguous/naturalistic stimuli. \\
Values (V) & Routed content & L4 input-related stream & Manipulating feedforward drive (e.g., L4-targeting) should change content available for routing but not necessarily the routing policy; content effects dissociate from contextual gating. \\
Queries (Q) & Select what to retrieve & L2/3 and L5 contextual stream & Context manipulations should modulate response gain/selectivity in L2/3 and L5 more strongly than L4; causal disruption of L2/3 or L5 should impair context-sensitive selection. \\
Keys (K) & Provide matchable addresses & L2/3 and L5 contextual stream via horizontal/feedback & Similarity structure between columns (functional connectivity) should be reshaped by feedback state; increased alignment during grouping/segmentation tasks. \\
Attention weights & Multiplicative gating & Gain modulation + dendritic nonlinearities & Evidence for multiplicative (not additive) modulation in relevant pyramidal compartments; task context changes effective synaptic impact without proportional change in presynaptic firing. \\
Self-attn connectivity & Defines interaction graph & Horizontal + feedback coupling & Task-dependent sparsification/rewiring of effective connectivity: same anatomy, different effective coupling; perturbing feedback should alter who influences whom more than raw firing rates. \\
Multi-head attention & Parallel routing policies & Subpopulations/dynamical modes within/across columns & Under rich context, subpopulations should show diverging routing profiles (different target influence patterns); diversity should shrink in impoverished stimuli and expand in naturalistic tasks. \\
FFN / MLP & Local recoding & Interlaminar transformation / local feature remap & Local nonlinear recoding should be measurable as a change of representational basis across laminae; disrupting specific interlaminar pathways should selectively impair feature recombination without removing contextual gating. \\
Residual (Add) & Preserve + update & Pooling/mixing with ongoing activity & Residual-like persistence predicts partial invariance of representations across short timescales despite strong contextual modulation; evidence of carry-over even when local recoding is strong. \\
LayerNorm (Norm) & Stabilize scale & Normalization-like gain control & Context-dependent normalization signatures (divisive-like) should correlate with stability of representational geometry; perturbing inhibitory gain control should destabilize coupling/selection. \\
Stack of blocks & Deep composition & Hierarchical modules with feedback & Higher-area perturbations should change lower-area effective coupling patterns (routing policy) rather than only adding bias; top-down should change selectivity structure earlier in time than expected from feedforward-only models. \\
\end{longtable}

\begin{longtable}{p{0.17\textwidth} p{0.22\textwidth} p{0.20\textwidth} p{0.33\textwidth}}
\caption{Relaxed assumptions with falsifiable consequences.}\label{tab:appendix-a2}\\
\toprule
\textbf{Theme} & \textbf{Relaxed assumption} & \textbf{What differs in cortex} & \textbf{Testable consequences / empirical handles} \\
\midrule
\endfirsthead
\toprule
\textbf{Theme} & \textbf{Relaxed assumption} & \textbf{What differs in cortex} & \textbf{Testable consequences / empirical handles} \\
\midrule
\endhead
\bottomrule
\endfoot
Literal implementation & Do not require explicit softmax/QKV matrices & Emergent gating and dynamics & Look for functional equivalence: context-dependent multiplicative effects and routing signatures, not literal computations. \\
Temporal unfolding & One-shot feedforward is not required & Iterative refinement & Time-resolved decoding should show progressive stabilization/refinement; later activity reflects feedback-shaped routing. \\
Top-down absence & Feedback is central & Bidirectional coupling & Opto/TMS/feedback perturbations should reshape effective connectivity and contextual modulation more than feedforward perturbations. \\
Token granularity & Tokens are not discrete & Continuous maps/columns & Representations should vary smoothly over cortical sheets; context can induce nonlocal coupling edges over that sheet. \\
Input dominance & Input may be weak relative to cortex & Context-dominated regimes & Under ambiguity, reducing thalamic drive may have smaller effects than disrupting intracortical loops on behavior/decoding outcomes. \\
Head separability & Heads are not explicit modules & Distributed subpopulations/modes & Identify head-like modes via clustering/latent dynamics; diversity should be task- and state-dependent. \\
Explicit attention matrix & No explicit attention matrix needed & Gating emerges & Measure routing with causal influence metrics (stimulation/Granger/transfer entropy/GLM coupling) rather than expecting explicit weight matrices. \\
One update per pass & Updates are iterative & Ongoing loops & Recurrent models should outperform feedforward models for time-course prediction; repeated passes (reverberation) should improve fit. \\
FFN locality & FFN is not a literal MLP & Laminar recoding + timing & Layer-specific perturbations should selectively disrupt feature recombination vs contextual selection, enabling dissociations. \\
\end{longtable}

\normalsize

Second, if laminar pathways implement something analogous to queries and keys, then context-dependent gain modulation should show characteristic laminar patterns. For example, manipulations of global or top-down context should preferentially modulate activity and synaptic efficacy in those laminae that carry feedback and horizontal connections (putatively “Q/K-like” pathways), while leaving the initial thalamic drive to the granular lamina (putatively “V-like” pathways) comparatively less affected. Such laminar-specific modulation would support the idea that context acts primarily by reshaping interaction structure rather than by uniformly scaling all inputs.

Third, if multi-head attention corresponds to multiple, partially independent routing schemes over the same inputs, one should observe distinct dynamical signatures associated with different heads---such as separable bands or phase relationships in oscillatory activity, or distinct classes of dendritic events on different branches---each selectively coupling particular sets of columns under appropriate task conditions. Failure to find any of these patterns---no increase in within-column diversity under context, no laminar-specific contextual gain, and no separable dynamical channels for routing---would constitute a serious challenge to the proposed mapping between transformer components and cortical microcircuits.

\subsection{What does the stacking achieve?}

So far, we have focused on the computations implemented by a single module---one cortical column together with its laminar circuitry---and how this can be understood as the analogue of a transformer block. In real cortex, however, such modules do not operate in isolation: they are repeated over topographic maps, connected horizontally, and, crucially, stacked into hierarchies with abundant feedback. It is therefore natural to ask what additional computations become possible when many of these transformer-like modules are arranged in depth and interact over time.

In the artificial transformer, stacking is, by design, purely feedforward within a single forward pass: each block receives the output embedding of the block below, transforms it, and passes the result upward, without any explicit top-down signal from higher to lower layers during that pass. By contrast, cortical hierarchies are bidirectionally connected. Higher areas send feedback projections to lower ones on distinct laminar pathways, and these projections can modulate ongoing processing rather than merely reading out its result. Our proposal can therefore be viewed as a biologically grounded extension of the transformer idea, in which the same kind of module participates in a recurrent hierarchy: over short time windows, activity in higher modules is fed back to lower modules and mixed into their recoding stage, thereby shaping which inputs are considered and how they are combined. This perspective is closely related in spirit to predictive-coding and other recurrent cortical models that emphasize top-down influence, but it casts those ideas in the more explicit language of queries, keys, values.

In the proposed architecture, stacking thus plays a central functional role rather than being a mere repetition of similar units. A useful way to understand this is by comparing with more “classical” recurrent networks and hierarchical cortical models \citep{Markov2014Anatomy}. In standard recurrent architectures, high-level representations can feed back to earlier processing stages, typically through additive modulation of activity; in biological terms, this corresponds to top-down projections that bias the firing rates or gain of neurons in lower areas according to the current context or prediction \citep{Lamme2000The}. Within the transformer-like scheme we propose for cortex, stacking goes a step further: top-down signals from higher modules do not only modulate the overall activity level of neurons in a lower module, but modulate the interaction structure within that lower module. When the output of an upper module is mixed into the recoding of the lower module’s output embedding, it effectively gates the multi-head attention operating there. In other words, the higher module shapes which queries and keys in the lower module are allowed to interact strongly, thereby altering which inputs are bound together and which relations are emphasized. Stacking thus creates a hierarchy in which high-level context does not simply turn lower-level neurons “up” or “down”, but reconfigures the pattern of connectivity by which lower-level information is combined, consistent with views of early visual cortex as a multiscale “blackboard” for cognitive coordination \citep{Roelfsema2016Early}.

This brings us back to the core mechanism of multi-head attention. As discussed above, the crucial innovation of the transformer relative to classical DNNs and CNNs is the multiplicative interaction of the form $(Q\cdot K^T)V$: a dot product between queries and keys determines how strongly each value contributes to the next representation \citep{Vaswani2017Attention}. In our cortical interpretation, this corresponds to a context-dependent gating of synaptic influence, where the match between a “query” pattern and a “key” pattern determines which inputs are selectively amplified or suppressed. Multi-head attention implements many such gating patterns in parallel, allowing different relational structures over the same input to be explored simultaneously. The stacking of modules then allows these gated interactions to be applied iteratively, with each higher module building more abstract context that feeds back to control the gating in lower modules. The question is how such a multiplicative, attention-like mechanism could be implemented biologically, and how such gating relates to more general cortical “buffering” or coordination mechanisms \citep{Phillips2015On}.

We suggest two complementary mechanisms: synchrony-based mechanisms and apical/basal two-point integration as candidate substrates for implementing attention-like gating, potentially co-existing in different laminae or timescales. The first possible implementation route invokes synchrony-based mechanisms. In this view, the match between query and key is realized by selective synchronization of the activity of neurons that carry compatible patterns, in line with the temporal correlation hypothesis and related accounts of binding by synchrony \citep{Singer1995Visual,Singer1999Neuronal}. Neurons whose recurrent and long-range connections encode a learned match between particular input features (the “keys”) and particular contextual demands (the “queries”) would tend to fire coherently when the match is strong. This coherence, in turn, makes the corresponding value inputs more effective at driving postsynaptic targets, consistent with theories of communication through coherence \citep{Fries2005A,Fries2015Rhythms}. In anatomical terms, the learned Q--K structure would be embedded in the tangential and feedback connectivity within and between columns. After learning, the pattern of who synchronizes with whom would be relatively fixed, so that the gating emerges from dynamic coordination rather than an explicit, separately represented “attention signal”. From the outside, such a circuit could look like an “attention-free transformer”: the selection of relevant inputs would be inherent in the connectivity and its oscillatory dynamics, not in an additional control pathway, consistent with empirical work linking oscillatory activity to long-range synchronization \citep{König1995Relation}.

The second, complementary implementation route emphasizes two-point integration by pyramidal neurons with distinct basal and apical dendritic compartments. Here, the basal dendrites of a neuron would receive the primary feedforward input, corresponding roughly to the values, while the apical tuft in upper layers would integrate contextual or top-down signals that encode queries or keys. A strong match between the pattern of activity arriving at the apical dendrite and the learned synaptic weights there could trigger a non-linear event (such as a calcium spike) that multiplicatively boosts the impact of coincident basal input on somatic firing \citep{Larkum1999A,Larkum2004Topdown,Major2013Active}. In this way, the neuron effectively computes a weighted multiplicative interaction: basal input is only fully expressed in the output when it is “gated in” by an appropriate apical context. Different neurons, or different dendritic branches within the same neuron, could then implement different attention-like “heads”, each tuned to a particular pattern of contextual modulation. Stacking of modules in this setting means that apical inputs from higher-order areas, carrying the results of more abstract computations, dynamically control which lower-level inputs are allowed to influence the ongoing representation, reproducing the layered, gated interactions that characterize multi-head attention in the transformer; related learning rules based on dendritic prediction provide a concrete candidate mechanism for how such apical gating could be trained \citep{Urbanczik2014Learning}.

\begin{table}[h]
\centering
\caption{Correspondence between transformer encoder/decoder operations,
cortical laminar compartments, and characteristic oscillatory signatures.
Cross-attention denotes the decoder querying the encoder representation
via L5 apical dendrites receiving L2/3 collateral input.}
\label{tab:oscillations}
\footnotesize
\setlength{\tabcolsep}{4pt}
\renewcommand{\arraystretch}{1.15}

\begin{tabularx}{\textwidth}{
>{\raggedright\arraybackslash}p{2.7cm}
>{\raggedright\arraybackslash}p{2.8cm}
>{\raggedright\arraybackslash}p{1.8cm}
>{\raggedright\arraybackslash}p{1.5cm}
>{\raggedright\arraybackslash}X
}
\toprule
\textbf{Transformer operation} &
\textbf{Cortical substrate} &
\textbf{Oscillation} &
\textbf{Band (Hz)} &
\textbf{Key references} \\
\midrule
Encoder self-attention (L2/3)
  & Supragranular pyramidal neurons
  & Gamma
  & 30--80
  & \cite{Bastos2012,Bastos2015,Bastos2020} \\

Encoder output / feedforward drive
  & L2/3 $\to$ higher areas
  & Gamma
  & 60--80
  & \cite{Bastos2015} \\

Decoder masked self-attention (L5)
  & Infragranular pyramidal neurons
  & Beta
  & 14--30
  & \cite{Bastos2012,Bastos2020} \\

Decoder feedback / top-down prediction
  & L5/6 $\to$ lower areas
  & Alpha/Beta
  & 8--28
  & \cite{Bastos2015,Bastos2020} \\

Decoder cross-attention (prediction*)
  & L5 apical dendrite $\leftarrow$ L2/3 collaterals
  & Gamma--Beta coupling
  & 30--80 / 14--30
  & \cite{Bastos2012,Bastos2020} \\

Thalamocortical input embedding (V)
  & Thalamus $\to$ L4
  & Theta / Gamma
  & 4--8 / 30--80
  & \cite{Bastos2015} \\
\bottomrule
\end{tabularx}
\end{table}

\subsection{Learning and plasticity}

To map the learning of a transformer block onto cortical columns, we must first confront the distinct computational objectives of the Value, Query, and Key pathways. In artificial networks \citep{Vaswani2017Attention}, the weight matrices $W_V$, $W_Q$, and $W_K$ are all updated uniformly via end-to-end backpropagation, yet their mathematical roles within the architecture differ fundamentally. The Value pathway is strictly linear with respect to the output, functioning as a feature extractor that provides the actual content to be processed. In contrast, the Queries and Keys dictate the dynamic routing of this content. They interact non-linearly via a dot-product operation followed by a competitive softmax function. Thus, the objective of queries and keys is not simply to extract sensory features, but to learn mutual alignment, ensuring that top-down/tangential contextual demands (Queries) correlate precisely with the appropriate bottom-up sensory evidence (Keys). Consequently, because the brain lacks a global backpropagation algorithm to implicitly parse these divergent mathematical goals---a biological implausibility long recognized in neuroscience \citep{Crick1989Recent,Lillicrap2020Backpropagation}---a biological transformer must rely on fundamentally distinct, localized plasticity rules: one tailored for extracting Value content, and a specialized, coincidence-based mechanism for aligning Queries and Keys.

Because the Value pathway functions mathematically as a linear content provider, its biological optimization can rely on well-established mechanisms of feedforward feature extraction. In the mammalian cortex, the primary bottom-up sensory drive, which propagates from the thalamus to lamina 4 and subsequently to the supragranular laminae, serves to construct a stable, descriptive dictionary of the external environment. Training the synaptic connections that encode these value representations does not require the specialized, multi-stream matching demanded by attention routing. Instead, the statistical regularities of continuous sensory input allow these feedforward synapses to be sculpted by standard Hebbian learning, most notably Spike-Timing-Dependent Plasticity (STDP) \citep{Markram1997Regulation,Bi1998Synaptic}. Furthermore, these representations are likely refined by local, self-supervised biological objectives, such as predictive coding, in which local laminar microcircuits continuously update their synaptic weights to minimize sensory prediction errors \citep{Rao1999Predictive,Bastos2012}. Ultimately, the biological instantiation of value learning does not present a novel mechanistic hurdle; it is naturally and sufficiently implemented by the brain’s classical, input-driven plasticity rules, which continuously build the robust repertoire of features that the attention mechanism will subsequently route.

While the linear Value pathway relies on standard feedforward feature learning, the non-linear alignment of Queries and Keys demands a dedicated biological coincidence detector capable of merging disparate anatomical streams. The neurophysiological hardware for such specialized detection is prominently realized in the thick-tufted Layer 5 pyramidal neurons of the neocortex. These neurons possess a highly compartmentalized morphology comprising two electrotonically segregated sites of synaptic integration: the proximal basal dendrites, which primarily receive bottom-up sensory drive (representing the Keys), and the distal apical tuft in Layer 1, which receives massive top-down feedback and contextual projections (representing the Queries) \citep{Körding2000Learning,Larkum2013A}. Extensive in vitro and in vivo studies have demonstrated that these two compartments are coupled by a powerful, non-linear active mechanism. While subthreshold apical input alone is highly attenuated and rarely influences somatic output \citep{Williams2002Dependence}, when apical depolarization coincides with basal somatic action potentials, it triggers a backpropagating action potential (BAC firing). This event unlocks voltage-gated calcium channels in the apical trunk, generating a prolonged dendritic calcium plateau potential that results in a high-frequency burst of somatic spikes \citep{Larkum1999A,Major2013Active}. The theoretical framework proposed by Körding and König \citep{Körding2000Learning,Körding2001Generation} demonstrates that this dual-integration morphology enables a powerful local learning rule: burst-dependent plasticity forces basal synapses to continuously predict the activity of the apical tuft. Translating this to our transformer mapping, the synapses forming the key pathway dynamically update to maximize their correlation with the top-down query pathway whenever a burst signals a match. Thus, the specialized compartmentalized morphology and burst-dependent plasticity of Layer 5 pyramidal neurons provide a fully localized, biologically plausible mechanism for continuously computing and aligning the dynamic Q−K attention matrix.

While early sensory areas leverage the "attention-free" efficiency of hard-wired connectivity, higher association areas must contend with non-stationary, context-dependent statistics that necessitate the active learning of Query (Q) and Key (K) alignments. In these regions, the mathematical objective of the supragranular circuit shifts from executing a fixed routing matrix to the dynamic refinement of the synaptic weights that dictate it. Because the brain lacks the global backpropagation required to synchronize these updates, we propose that this alignment is achieved through a localized, coincidence-based plasticity rule.

To fully realize a transformer-like attention mechanism, computing the dot-product similarity between Queries and Keys is necessary but not yet sufficient; the resulting matches must also be subjected to a highly non-linear, competitive normalization, mathematically defined as the Softmax function. In artificial transformers, the Softmax operation ensures that attention weights sum to one, meaning that as a network attends more to one specific feature or token, it must mathematically suppress its attention to others. Biologically, the global, instantaneous calculation of a Softmax function across an entire sequence is unclear. However, the cerebral cortex naturally implements a powerful approximation of this competitive routing through local lateral inhibition and divisive normalization \citep{Reynolds2009The}. When a specific subset of Layer 5 pyramidal neurons detects a strong Q - K match and fires a burst of action potentials, these neurons recruit local inhibitory interneurons, such as parvalbumin-positive basket cells and somatostatin-positive cells \citep{Kapfer2007Supralinear,Silberberg2007Disynaptic}. This recurrent inhibitory drive swiftly suppresses the activity of neighboring, less-activated pyramidal neurons within and across adjacent columns. Consequently, this local inhibitory microcircuit creates a "winner-take-all" dynamic that functionally mirrors the exponentiation and divisive suppression characteristic of the Softmax operation. Thus, while the excitatory apical-basal coincidence computes the raw attention score, it is the dense local inhibitory network that enforces the necessary competitive normalization, ensuring that only the most highly relevant representations are routed forward.

Artificial networks rely on the mathematically uniform, yet biologically implausible, algorithm of global backpropagation to simultaneously sculpt feature extraction, dynamic routing, and parallel attention. In contrast, the cortex appears to achieve the exact same computational motif - context-dependent, gated routing over shared inputs - by physically segregating these operations across distinct anatomical and biophysical substrates. By assigning the linear generation of Values to standard feedforward pathways shaped by classical Hebbian plasticity, and the non-linear alignment of Queries and Keys to the specialized, coincidence-detecting compartments of Layer 5 pyramidal neurons, the column bypasses the need for end-to-end gradient descent. Furthermore, by utilizing local inhibitory microcircuits to enforce competitive normalization (Softmax) and leveraging the natural variability of non-classical receptive fields to compute parallel relations (Multi-Head Attention), the column maximizes its computational density. In this view, the intricate, layered complexity of the cortical column is not merely a biological quirk, but an elegant, fully localized evolutionary solution to the very same routing and binding problems solved by the modern transformer block.

\section{Extensions}

Mapping the transformer architecture onto the laminated structure of the brain may cast a new light on several properties of biological neural networks that are currently underused in mainstream modeling \ref{tab1:transformer_cortex_mapping}. In our proposal, the laminar circuitry of cortex is no longer treated as a largely decorative anatomical detail, but as the key substrate for implementing transformer-like computations, building on canonical descriptions of neocortical circuitry \citep{Douglas2004Neuronal,Harris2015The}. This perspective forces us to ask more specific questions: What exactly does recurrent connectivity contribute when embedded in a laminar, attention-like module? How might spike timing, oscillations, and neuromodulatory systems shape or implement multi-head attention? And what roles might subcortical structures play in gating and routing information in a transformer-like brain architecture? In this final section, we briefly outline some of these directions.

\subsection{Recurrent connectivity and Temporal Scaffolding}

Recurrent connectivity is already a major research topic in computational neuroscience and machine learning, with work spanning dynamical systems, attractor networks, working memory, and models of reaction times and decision dynamics \citep{Wang2002Probabilis,Wong2006A,Rabinovich2008Transient}. In many of these studies, recurrence is treated in fairly generic terms: recurrent connections are added, and one asks what behaviors emerge or how they alter fixed points and trajectories in state space. In the framework proposed here, recurrence is embedded inside a transformer-like module with explicit roles for queries, keys, and values mapped onto laminae and cell types. This raises sharper questions: which recurrent loops implement something like key--query interactions, and which loops primarily serve to sustain or stabilize value representations over time? How do different forms of recurrence---within layers, across layers, and between modules---affect the effective depth and capacity of the cortical “transformer stack”? Answering these questions could connect the extensive literature on recurrent cortical dynamics more directly to concrete architectural motifs familiar from artificial networks, and it connects naturally to evidence that recurrence is required to capture the representational dynamics of the human visual system \citep{Kietzmann2019Recurrence}.

A fundamental challenge in mapping transformer operations onto the brain is reconciling the architecture’s requirement for discrete, causal sequences with the cortex’s continuous, recurrent dynamics. We propose that the brain resolves this through a hierarchical temporal scaffolding: cortical circuits use rhythmic synchronization to discretize information into functionally atomic units. Building on the Communication Through Coherence (CTC) framework, gamma-band oscillations (30–90 Hz) organize neural firing into brief excitatory windows, creating the high-resolution "packets" necessary for feature representation. When nested within slower theta rhythms (5–8 Hz), these gamma cycles provide a biological analogue to the transformer’s sequential token processing—multiplexing discrete representational items within a single indexing frame. This temporal discretization is physically instantiated across the cortical column through a distinct spectral-laminar motif. We identify the supragranular laminae (L2/3) as the biological Encoder, characterized by prominent gamma-band activity. Here, L2/3 pyramidal neurons integrate thalamocortical "Values" from L4 with top-down "Queries" (L2/3) and lateral "Keys" (horizontal L2/3 axons) to generate a feedforward encoded representation. This mapping is supported by electrophysiological evidence showing that gamma oscillations preferentially carry feedforward signals across the visual hierarchy, representing the initial encoding of sensory evidence. Conversely, we identify the infragranular layers (L5) as the biological Decoder, operating in the alpha/beta regime (8–28 Hz). These layers receive the encoded L2/3 representation as input for a cross-attention operation, while simultaneously managing the autoregressive self-attention of the decoder via thalamocortical loops (e.g., L5 to pulvinar). This spectral segregation—gamma for encoding/feedforward and alpha-beta for decoding/feedback—provides the necessary physical and temporal separation to prevent signal interference between the "what is there" (encoder) and the "what is predicted" (decoder). By mapping the transformer core onto this anatomy, we suggest that the cortical column does not just process information; it physically segregates the encoder and decoder into distinct laminar compartments that communicate via characteristic frequencies tailored to their specific computational roles.

\subsection{Spiking networks}

Spiking networks, spike-timing-dependent plasticity (STDP), causal learning rules, and oscillations provide another rich set of mechanisms that our proposal has so far only touched upon \citep{Bi1998Synaptic,Caporale2008Spike,Buzsaki2006Rhythms}. One intriguing possibility is that “attention-free transformers” might be implemented in early sensory areas, especially the lower visual hierarchy, where the statistics of natural images could be hard-wired into the pattern of tangential connectivity via activity-dependent plasticity. In such a scenario, the effective Q--K structure would be encoded in the long-range horizontal connections whose strength and specificity reflect the co-occurrence statistics of contour segments, textures, or motion patterns in natural scenes, consistent with work on the functional specificity of local synaptic connections \citep{Ko2011Functional}. Oscillations and synchrony would then provide the fast dynamics necessary to instantiate multi-head attention: distinct frequency bands, phase relationships, or cell assemblies could implement different heads operating in parallel over the same underlying inputs, in line with theories of communication through coherence \citep{Fries2015Rhythms}. In this view, the multi-head structure of attention in artificial transformers corresponds to a repertoire of distinct, partially overlapping synchrony patterns that route information along different subsets of the tangential network. More broadly, normative accounts of spontaneous cortical activity and naturalistic stimulus statistics suggest how such connectivity and dynamics could be shaped by experience \citep{Berkes2011Spontaneou,Betsch2004The}.

\subsection{Different time scales}

A third domain where the transformer perspective may be fruitful concerns transmitter systems and time constants. It is often assumed, both in theory and in data-driven work, that effective time constants of processing increase along cortical hierarchies, such that higher areas integrate information over longer windows than lower ones \citep{Murray2014A,Chaudhuri2015A}. This has been formalized in models with explicitly imposed time-scale hierarchies, and in normative frameworks like slow feature analysis \citep{Wiskott2002Slow}, as well as in proposals that synaptic or intrinsic properties adapt to match relevant temporal statistics (e.g., temporally stable representations and local memory in ventral stream models) \citep{Wyss2003A}. At the biophysical level, this could be implemented through a rich mixture of ionotropic and metabotropic receptors, co-release of fast and slow transmitters, and diverse short- and long-term synaptic dynamics, consistent with broad accounts of neuromodulatory control of circuit dynamics \citep{Marder2012Neuromodul}. From the viewpoint of transformers, these temporal aspects acquire an additional dimension: attention mechanisms in artificial models suffer from quadratic scaling with context length, and considerable effort has gone into designing sparse, low-rank, or otherwise more efficient attention variants to handle longer sequences \citep{Child2019Generating,Kitaev2020Reformer,Beltagy2020Longformer}. Biological circuits face an analogous challenge in integrating information over behavioral timescales with limited resources. It is tempting to speculate that the diversity of neuromodulatory systems, receptor kinetics, and synaptic plasticity rules implements a kind of “temporal attention,” selectively maintaining and highlighting information across time in a way that parallels how transformers allocate attention across tokens, and it may relate to more general principles of Bayesian integration in sensorimotor learning \citep{Kording2004Bayesian}.

\subsection{Subcortical structures}

Further, subcortical structures deserve a more central place in any brain--transformer analogy. The basal ganglia, for instance, have long been discussed as a system for gating actions and controlling which cortical representations are expressed or updated, both in computational models of action selection and in cortico--basal-ganglia working-memory frameworks \citep{Gurney2001A,OReilly2006Making}. This has an obvious similarity to the gating role played by attention in transformers, where only selected inputs are allowed to influence the current computation strongly. A transformer-informed view might sharpen this analogy by asking whether basal ganglia circuits implement something closer to value gating (deciding which outputs are propagated), query gating (controlling which contexts are considered), or key gating (selecting which stored patterns can be matched). Furthermore, comparative anatomical data suggest systematic differences in the composition of striatal circuits between humans and other animals, including changes in the proportion and diversity of interneurons. Such quantitative changes could support more fine-grained or flexible gating in human cortico--striatal loops, potentially expanding the effective “vocabulary” of attention-like operations. Exploring these possibilities could help bridge current gaps between systems neuroscience, theories of cognitive control, and the architectural principles emerging from transformer models, while staying grounded in established functional neuroanatomy and dopamine-based computational accounts \citep{Lanciego2012Functional,Frank2005Dynamic}.

\section{Alternative approaches}

A small but growing body of work has begun to ask how transformers, and in particular self-attention, might be implemented in more biologically grounded circuits. One influential line focuses on neuron--astrocyte networks: Kozachkov and colleagues show that tripartite synapses can, in principle, implement the core components of self-attention, with astrocytic dynamics providing the normalization and multiplicative gating that parallel the softmax attention operator \citep{Kozachkov2023Building}. They extend this framework to dense associative memory models in which neuron--astrocyte dynamics converge to attention-like fixed points over stored patterns \citep{Kozachkov2025Neuronastr,Kozachkov2023Neuronastr}. A second, largely orthogonal line develops spiking and neuromorphic transformers, such as Spikformer and related architectures, which replace rate units with spiking neurons and implement attention using spike-based Q, K, and V representations, sometimes coupled to local plasticity rules so that effective attention weights emerge from synaptic changes rather than explicit matrix multiplications \citep{Zhou2023Spikformer,Mondal2025Attention}. These approaches demonstrate that transformer-style computations can be realized in biophysically flavored systems---either at the level of detailed synaptic and glial dynamics or at the level of spike-based algorithms aimed at energy-efficient hardware---but they remain relatively agnostic about the mesoscopic architecture of the cortex. In contrast, our proposal takes the laminar cortical column as the primary unit of analysis and asks whether its canonical microcircuitry can be interpreted as a reusable transformer-like module: queries, keys, and values are assigned to specific laminar pathways and recurrent loops; multi-head attention is mapped onto parallel neuronal subpopulations within a column; and self-attention across “tokens” is realized by horizontal and feedback connectivity between columns. As a consequence, the experimental predictions of these frameworks differ sharply. Neuron--astrocyte accounts imply that disrupting astrocytic signaling or tripartite synapse dynamics should directly impair attention-like normalization, whereas our account predicts strong task- and context-dependent differentiation of responses across laminae and within columns, even when classical receptive fields appear similar under simple stimuli. Likewise, spiking transformer models make distinctive claims about the role of precise spike timing and specific spike-based plasticity rules in implementing attention, whereas our proposal emphasizes laminar patterns of gain modulation and long-range grouping. Taken together, these complementary but competing pictures suggest a rich experimental agenda: by systematically probing spike timing, glial contributions, and laminar-specific contextual effects under naturalistic tasks, it should be possible not only to test whether the cortex implements an attention-like computational motif but also to discriminate between alternative biological implementations of transformer-style computation.

The proposed mapping of transformer-like motifs onto the cortical column finds significant structural and functional alignment with the recently articulated Lamina 6b Theory of Attention (LAT). While our framework identifies lamina 5 (L5) pyramidal neurons as the primary biophysical substrate for Query-Key (QK) alignment via Backpropagating Action Potential-activated Calcium (BAC) firing, the LAT provides a necessary regulatory layer by positioning lamina 6b (L6b) as the "conductor" of these attention loops. In the context of a biological transformer, L6b likely functions not as the source of the specific feature-based "Key" or "Query" signals—which remain the domain of local and tangential intracortical connectivity—but as an extrinsic "attention gate" or "temperature" modulator. By leveraging its unique, facilitating projections to the L5 apical tuft and its reciprocal feedback loops with the higher-order thalamus (HoT), L6b may dynamically set the gain for dendritic coincidence detection. Mathematically, this suggests that L6b acts as a control parameter for the softmax-like competitive normalization within the column, effectively scaling the "sharpness" of the attention matrix in response to global arousal or task-demand signals, such as those mediated by orexinergic inputs. This synthesis suggests a hierarchical control scheme where local transformer blocks perform high-resolution feature routing, while the L6b-thalamic system provides the state-dependent stability required to maintain these representations across longer temporal windows. Integrating L6b into the transformer-cortex mapping thus reconciles the high-speed computational requirements of multiplicative routing with the slower, more stable dynamics of cognitive focus and arousal.

\section{Conclusion}

In this essay, we have put forward the specific hypothesis the mammalian brain implement a transformer-like architecture through the lamina structure of the cerebral cortex. By spelling out this proposal and working through a number of concrete mappings, we have inevitably made assignments of function and interpretation that are, at least in part, wrong or oversimplified. The correspondence between cortical laminae, recurrent loops, and transformer components is almost certainly more complex and heterogeneous than our schematic diagrams suggest. Nonetheless, we hope that the core idea---that cortical microcircuits implement context-dependent, gated interactions over shared inputs in a way that is structurally analogous to transformers---offers a useful interpretive framework for thinking about signal flow and computation in the brain. At a minimum, it highlights a set of questions that can be addressed empirically and computationally. More optimistically, we hope that this perspective will help to revive and deepen the exchange between research on artificial neural networks and systems neuroscience, encouraging both communities to revisit the laminated cortex not just as anatomy, but as a candidate implementation of a powerful, transformer-like computational motif.

\section{Acknowledgments}

Mario would like to acknowledge insightful discussions of parts of this idea with Jorge Mejias, Lennart Landsmeer, Christos Styridis, Tom Viering, Amirreza Mohavedin, Egidio D'Angelo, and Michele Giugliano.

\section{AI use statement}
All ideas, interpretations, and conceptual developments in this manuscript are those of the authors. AI tools were used only to support discussion of ideas, language editing, research copyediting, and paragraph structuring. The authors take full responsibility for the content of the manuscript.

\bibliographystyle{abbrvnat}
\bibliography{references}

\end{document}